\documentclass[twocolumn,showpacs,preprintnumbers,amsmath,amssymb]{revtex4}
\usepackage{graphics}
\usepackage{epsfig}
\usepackage{bm}

\begin{document}
\title{Effect of Co and Fe on the inverse magnetocaloric properties of Ni-Mn-Sn}
\author{Thorsten Krenke$^{a)}$\footnotetext{$^{a)}$Present address: ThyssenKrupp Electrical Steel, F\&E Ge, Kurt-Schumacher-Str. 95, D-45881 Gelsenkirchen, Germany
}, Ey\"{u}p Duman$^{b)}$\footnotetext{$^{b)}$Present address:
Heinrich Heine Universit\"{a}t D\"{u}sseldorf, Institut f\"{u}r
Physik der kondensierten Materie, Abteilung f\"{u}r
Materialwissenschaft, D-40225 D\"{u}sseldorf Germany}, Mehmet
Acet}

\affiliation{Fachbereich Physik, Experimentalphysik,
Universit\"{a}t Duisburg-Essen, D-47048 Duisburg, Germany}

\author{Xavier Moya, Llu\'{i}s Ma\~nosa, Antoni Planes}
\affiliation{Facultat de F\'isica, Departament d'Estructura i
Constituents de la Mat\`eria, Universitat de Barcelona, Diagonal
647, E-08028 Barcelona, Catalonia, Spain}
\date{\today}
\begin{abstract}
At certain compositions Ni-Mn-$X$ Heusler alloys ($X$: group
IIIA-VA elements) undergo martensitic transformations, and many of
them exhibit inverse magnetocaloric effects. In alloys where $X$
is Sn, the isothermal entropy change is largest among the Heusler
alloys, particularly in Ni$_{50}$Mn$_{37}$Sn$_{13}$ where it
reaches a value of 20 Jkg$^{-1}$K$^{-1}$ for a field of 5T. We
substitute Ni with Fe and Co in this alloy, each in amounts of 1
at\% and 3 at\% to perturb the electronic concentration and
examine the resulting changes in the magnetocaloric properties.
Increasing both Fe and Co concentrations causes the martensitic
transition temperature to decrease, whereby the substitution by Co
at both compositions or substituting 1 at\% Fe leads to a decrease
in the magnetocaloric effect. On the other hand, the
magnetocaloric effect in the alloy with 3 at\% Fe leads to an
increase in the value of the entropy change to about 30
Jkg$^{-1}$K$^{-1}$ at 5T.
\end{abstract}


\maketitle

\section{Introduction}

Since the observation of martensitic transformations in Ni-Mn
based Heusler alloys \cite{Webster84}, and the later discovery of
the magnetic shape memory effect (MSM) in Ni-Mn-Ga
\cite{Ullakko96}, there has been growing interest in the
magneto-mechanical properties of Ni-Mn-$X$ systems ($X$: group
IIIA-VA elements)
\cite{Acet02,Sutou04,Krenke05long,Krenke06long,Oikawa06,Koyama06}.
Martensitic transformations in these systems are found roughly in
a valence electron concentration range of $7.4\leq e/a\leq 8.5$
electrons per atom. Next to the MSM effect, large magnetocaloric
effects (MCE) have been reported in the vicinity of the
martensitic transformation in Ni-Mn-Ga \cite{Marcos0203} and other
Ni-Mn-based Heusler alloys such as Ni-Mn-Sn and Ni-Mn-In
\cite{Krenke05a, Krenke06b, Han06}. The MCE in these systems
relies on the large and rapid temperature variation in the
magnetic interaction around the martensitic transition.
Furthermore, the MCE in Ni-Mn-based Heuslers is usually "inverse,"
i.e., the alloy cools on applying a magnetic field adiabatically
instead of heating \cite{Krenke05a}. The magnitude of the inverse
MCE is about 20 J kg$^{-1}$K$^{-1}$ for 5T in
Ni$_{50}$Mn$_{37}$Sn$_{13}$ being comparable with those of giant
MCE material \cite{Gschneidner00,Gschneidner05,Bruck05}. The use
of normal and inverse MCE materials in the form of composites can
affect the heat transfer efficiency favorably in heating or
cooling units based on the MCE.

To understand the influence of introducing small amounts of
transition elements on the size of the MCE in
Ni$_{50}$Mn$_{37}$Sn$_{13}$, we have investigated the entropy
change around the martensitic transition of
(Ni,Co)$_{50}$Mn$_{37}$Sn$_{13}$ and
(Ni,Fe)$_{50}$Mn$_{37}$Sn$_{13}$, in which Ni was replaced by Fe
and Co, each in amounts of 1 at\% and 3 at\%. Higher
concentrations of Fe and Co lead nearly to the complete
suppression of the martensitic state. The entropy change was
determined from field dependent magnetization measurements.

\section{Experimental}

Ingots of about 3 g were prepared by arc melting pure metals under
argon atmosphere in a water cooled Cu crucible. The ingots were
then encapsulated under argon atmosphere in quartz glass and
annealed at 1273 K for 2 hours. Afterwards, they were quenched in
ice-water. The compositions of the alloys were determined by
energy dispersive x-ray photoluminescence analysis (EDX) and are
given in at\% in Table \ref{tab:Composition}. The valence electron
concentrations $e/a$ are also given in the table. This is
calculated as the concentration weighted sum of the number of 3$d$
and 4$s$ electrons of Ni and Mn and the number of 4$s$ and 4$p$
electrons of Sn. An estimated error of $\pm0.1$\% in determining
the concentration leads to an error of $\pm0.007$ in the value of
$e/a$.

\begin{table} \caption{\label{tab:Composition} Compositions of the samples in at\% determined by EDX analysis and the valence electron concentrations per atom $e/a$. The sample containing no Fe or Co is referred to as 'Reference'.}
\begin{ruledtabular}
\begin{tabular}{lcccccc}
Sample & Ni & Co & Fe & Mn & Sn & $e/a$\\
\hline
Reference & 49.9 & - & - & 37.1 & 13.0 & 8.107\\
Co1 & 48.7 & 1.2 & - & 36.9 & 13.2 & 8.089\\
Co3 & 47.0 & 3.1 & - & 36.6 & 13.3 & 8.073\\
Fe1 & 49.1 & - & 0.9 & 36.7 & 13.3 & 8.083\\
Fe3 & 46.9 & - & 3.0 & 36.8 & 13.3 & 8.038\\
\end{tabular}
\end{ruledtabular}
\end{table}

Polycrystalline samples were cut from the ingots using a low speed
diamond saw and used as samples for magnetization and calorimetric
studies. For differential scanning calorimetry (DSC) measurements,
one side of the samples were ground with 1200 grid SiC abrasive to
insure proper thermal contact. Calorimetric measurements were
carried out in the temperature range 100 K$ \leq T\leq$ 350 K.
Typical heating and cooling rates were 0.5 K min$^{-1}$.

The temperature dependence of the magnetization $M(T)$ was
measured in magnetic fields $\mu_0H$ of 5 mT and 5 T in the
temperature range 5 K $\leq T\leq380$ K using a superconducting
quantum interference device magnetometer. Prior to the
measurements, the samples were prepared in a zero-field-cooled
state (ZFC) by cooling it from 380 K to 5 K in the absence of a
magnetic field. Subsequently, an external field was applied, and
the data were taken on increasing temperature up to 380 K. Then,
without removing the external field, the data were taken on
decreasing temperature, giving the field-cooled (FC) branch. As a
last step, again without removing the external field, the
magnetization was measured on increasing temperature. The last
step is denoted as the field-heated (FH) sequence. Any hysteresis
in the FC and FH sequences is expected to be associated with a
structural transition, whereas any splitting of the ZFC and FH
curves is expected to be associated with pinning due to
antiferromagnetic (AF) or noncollinear magnetic structures
existing within the ferromagnetic (FM) matrix. Such structures can
reside within twin boundaries if the system is martensitic.

\section{Results an discussion}

\subsection{Calorimetric studies}

The results of the calorimetric experiments are collected in Fig.
1. Next to the features associated with the martensite start
$M_s$, martensite finish  $M_f$, austenite start $A_s$, and
austenite finish $A_f$ temperatures, features associated with the
Curie temperature of the austenitic state $T_C^A$ are also
observed. For Co1, $T_C^A$ is observed only in the cooling curve.
In the heating curve, such a feature is masked by the intense peak
associated with the reverse transition. The positions of $T_C^A$
are in good agreement with those determined from the magnetization
measurements (section \ref{MM}). The structural transition
temperatures determined from the calorimetry data are indicate
with a DSC subscript and are collected in Table \ref{tab:Temps}.
Values for the reference sample Ni$_{50}$Mn$_{37}$Sn$_{13}$ from
earlier data are also included in the table \cite{Krenke05a}. The
transition temperatures decrease with respect to those of the
reference sample as Fe or Co is substituted for Ni due to the
decrease in the valence electron concentration.

\begin{table}
\caption{\label{tab:Temps} Martensite start and finish
temperatures ($M_s, M_f$), austenite start and finish temperatures
($A_s, A_f$) determined by DSC and $M(T)$ measurements ($DSC$ and
$M$ in superscript). $T_C^A$ are also given. The reference sample
containing no Fe or Co is labeled as 'Ref'.}
\begin{ruledtabular}
\begin{tabular}{lcccccccccc}
Sample & $M_s^{DSC}$ & $M_s^{M}$ & $M_f^{DSC}$ & $M_f^{M}$ & $A_s^{DSC}$ & $A_s^{M}$ & $A_f^{DSC}$ & $A_f^{M}$ & $T_C^A$\\
$ $ & (K) & (K) & (K) & (K) & (K) & (K) & (K) & (K) & (K)\\
\hline\\
Ref & 307 & 303 & 289 & 265 & 295 & 275 & 318 & 309 & 311\\
Co1 & 306 & 305 & 257 & 259 & 277 & 267 & 325 & $-$ & 316\\
Co3 & 300 & 294 & 217 & 205 & 229 & 214 & 303 & 323 & 335\\
Fe1 & 259 & 261 & 209 & 208 & 228 & 221 & 279 & 280 & 316\\
Fe3 & 187 & 195 & 156 & 141 & 177 & 150 & 216 & 196 & 315\\
\end{tabular}
\end{ruledtabular}
\end{table}

\begin{figure}
\includegraphics[width=8cm]{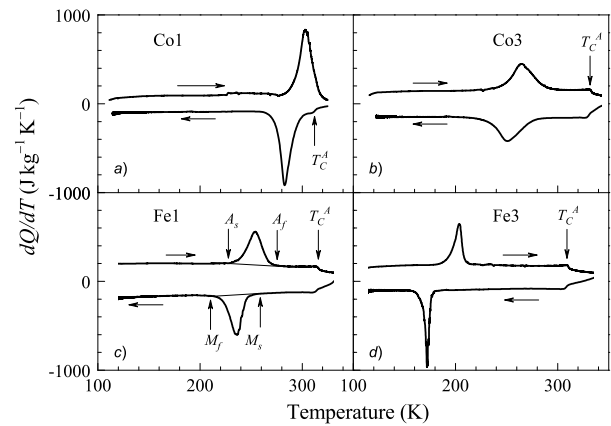}
\caption{\label{calorimetry} $dQ/dT$ versus temperature for the
alloys undergoing martensitic transformations (a) Co1, (b) Co3,
(c) Fe1, and (d) Fe3. Horizontal arrows indicate direction of
temperature change. Vertical arrows in part (b) show as an example
the positions of the martensitic and austenitic transition
temperatures.}
\end{figure}

\subsection{Magnetization measurements}\label{MM}

\begin{figure}
\includegraphics[width=7cm]{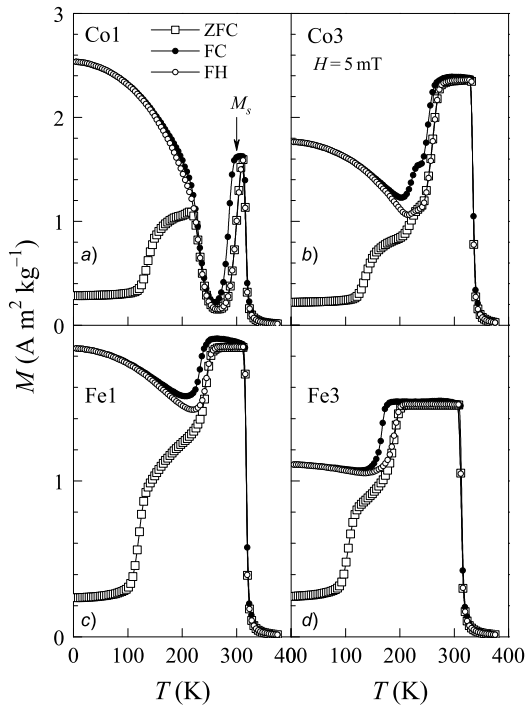}
\caption{\label{MT5mT} ZFC, FC, and FH $M(T)$ in 5 mT of a) Co1,
b) Co3, c) Fe1, and d) Fe3.}
\end{figure}

The temperature dependence of the FC, ZFC, and FH magnetization
$M(T)$ measured in 5 mT is shown in Fig. \ref{MT5mT}. All samples
order ferromagnetically in the austenitic state below $T_C^A$.
With respect to $T_C^A$=311 K of the reference alloy, $T_C^A$
increases with increasing Fe and Co concentrations, whereby the
increase is more significant for the case of Co (Tab.
\ref{tab:Temps}). For $T<T_C^A$, $M(T)$ is at the demagnetization
limit in 5 mT and remains essentially temperature insensitive in
all samples until it begins to drop at $M_s$. The $M_s^{M}$ values
given in Table \ref{tab:Temps} are taken as the temperature where
the maximum in the FC-$M(T)$ data is observed. These values are in
good agreement with $M_s^{DSC}$, whereby only the Fe3 sample has a
larger difference in the values of $M_s$ determined by the two
methods. At lower temperatures, $M(T)$ runs through a minimum for
all samples. The temperatures related to the minima lie somewhat
lower than $M_f$ determined from the DSC measurements. Similarly,
$A_s$ and $A_f$ lie around the temperatures at which the FH-$M(T)$
curves exhibit a minimum and a maximum respectively. Since any of
these temperatures vary rapidly with composition, any major
differences of the transition temperatures determined by the two
methods can be related to small compositional differences in the
samples used for the $M(T)$ and for the DSC experiments. In the
case of the Co3 sample, an additional feature within the
transition region is observed, which is indicative of an
intermartensitic transition. However, the data are not conclusive,
and further evidence is required to explain this feature.

\begin{figure}
\includegraphics[width=7cm]{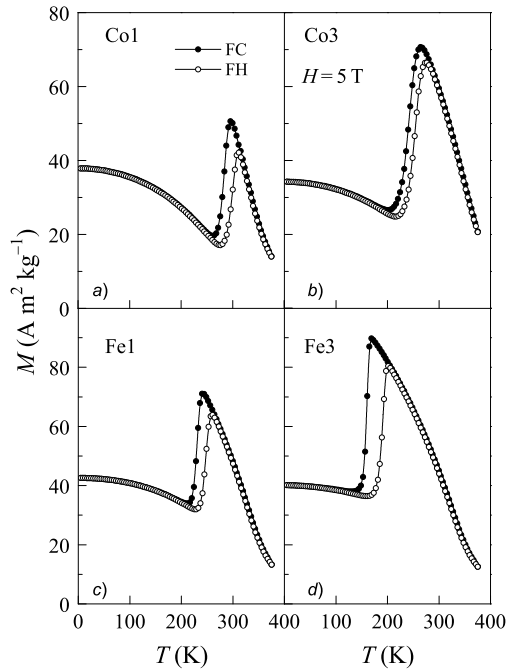}
\caption{\label{MT5T} (a) ZFC, FC, and FH $M(T)$ in 5 T of a) Co1,
b) Co3, c) Fe1, and d) Fe3.}
\end{figure}

The ZFC data in Fig. 2 begin at a low magnetization value of about
0.2 Am$^2$kg$^{-1}$ due to the essentially random spatial
configuration that the moments acquire while cooling through $T_C$
down to low temperatures. At low temperatures, the spin system is
in a frozen state. On heating, $M(T)$ remains constant up to about
100 K in all samples. At this temperature, the thermal energy
begins to overcome the exchange anisotropy of the frozen state in
the 5 mT measuring field, and the magnetization begins to
increase. At a slightly higher temperature, the rate of increase
slows down, and eventually, the ZFC and FH curves merge. This
occurs at a temperature between $A_s$ and $A_f$ as the proportion
of the austenitic phase increases, and along with it,
ferromagnetic exchange gains strength.

$M(T)$ measured in 5 T is shown in Fig. 3. A difference in the ZFC
and FH $M(T)$ curves no longer exist in this field, and ZFC data
are omitted. However, a hysteresis is still present in the FC and
FH curves for all samples. The hysteresis narrows with increasing
Co content, while it becomes broader with increasing Fe content.
The temperature corresponding to the peak in the FC-$M(T)$ curve
shifts to lower values with respect to $M_s$ found from the data
in Fig. 2. The shift is about 10 K for the Co1 sample and 20-30 K
for the other samples. For all samples, the magnetization drops as
the structure transforms from austenite to martensite. This drop
can be related to the fact that in Mn-based Heusler systems, the
exchange interaction strongly depends on the Mn-Mn distance, and
any change in the distance caused by a change in the crystal
structure can modify the interactions and introduce
antiferromagnetic exchange. This has been shown to be the case for
a Ni$_{50}$Mn$_{36}$Sn$_{14}$ alloy, in which short range
antiferromagnetic exchange was found to be present between Mn
atoms located at the 4(b) positions of the austenite phase and
that the exchange strengthens in the martensitic state
\cite{Brown06}.

\subsection{Magnetocaloric effects}\label{sectMCE}

To determine the size of the magnetocaloric effect (MCE) in the
vicinity of the transition we use the Maxwell relation

\begin{equation}
\Delta S(T,H) = \mu_0\int^{H}_{0}\left(\frac{\partial M}{\partial
T}\right)_{H} dH,
\end{equation}from which the MCE can be estimated numerically from $M(H)$ isotherms.
The $M(H)$ curves for the samples in the temperature range of
their respective structural transition temperatures are given in
Figs. 4 and 5. The curves are obtained on increasing field and
decreasing temperature as indicated by the arrows. The sequence of
measurements shown with open circles (red) shows an increase in
$M(H)$ with decreasing temperature down to $M_s$ for all samples.
The behavior reverses below $M_s$. For Fe1, Fe3, and Co1, $M(H)$
increases steadily up to 5 T without any major feature. For Co3,
there is a tendency for a metamagnetic-like transition for
temperatures below $M_s$ as can be seen from the development of an
upturn of the data with increasing field at an inflection point
around 2$-$3 T. This feature is expected to be due to the onset of
a field induced transition from the martensitic to the austenitic
state, which is the state of higher magnetization. The
observability of this effect only in the Co3 sample is because,
the shift in $M_s$ to lower temperatures in the applied fields in
these experiments exceeds the width of the hysteresis in $M(T)$
(Fig. 2). For the other samples, the hysteresis is too broad to
observe a field induced transition up to 5 T.

\begin{figure}
\includegraphics[width=8cm]{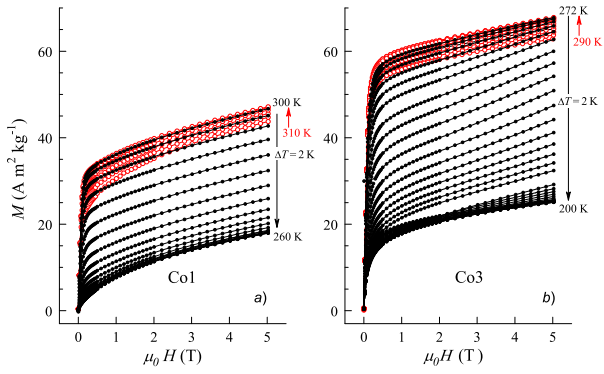}
\caption{\label{MH1} (color online) The field dependence of the
magnetization for a) Co1 and  b) Co3. The open circles refer to
data for which the magnetization decreases from 300 K to 310 K for
Co1 and from 272 K to 290 K for Co3. The vertical arrows indicate
the temperature sequence of data collection.}
\end{figure}

\begin{figure}
\includegraphics[width=8cm]{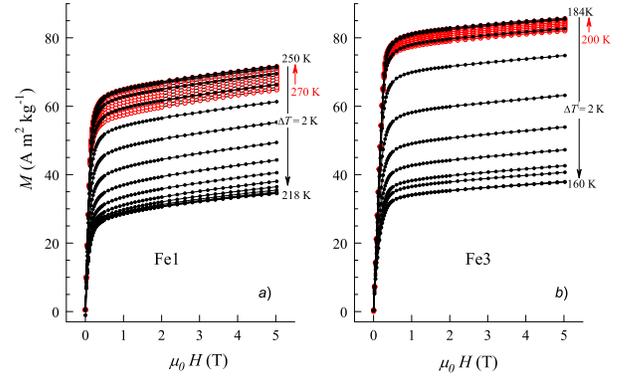}
\caption{\label{MH2} (color online) The field dependence of the
magnetization for a) Fe1 and  b) Fe3. The open circles (red) refer
to data for which the magnetization decreases from 250 K to 270 K
for Fe1 and from 184 K to 200 K for Fe3. The vertical arrows
indicate the temperature sequence of data collection.}
\end{figure}

\begin{figure}
\includegraphics[width=8cm]{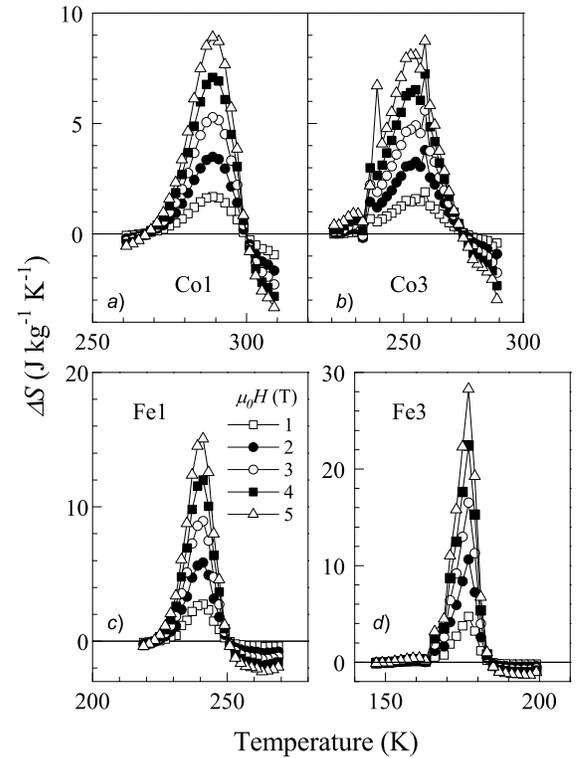}
\caption{\label{MCE} The entropy change for a) Co1, b) Co2, c)
Fe1, and d) Fe2.}
\end{figure}

\begin{figure}
\includegraphics[width=8cm]{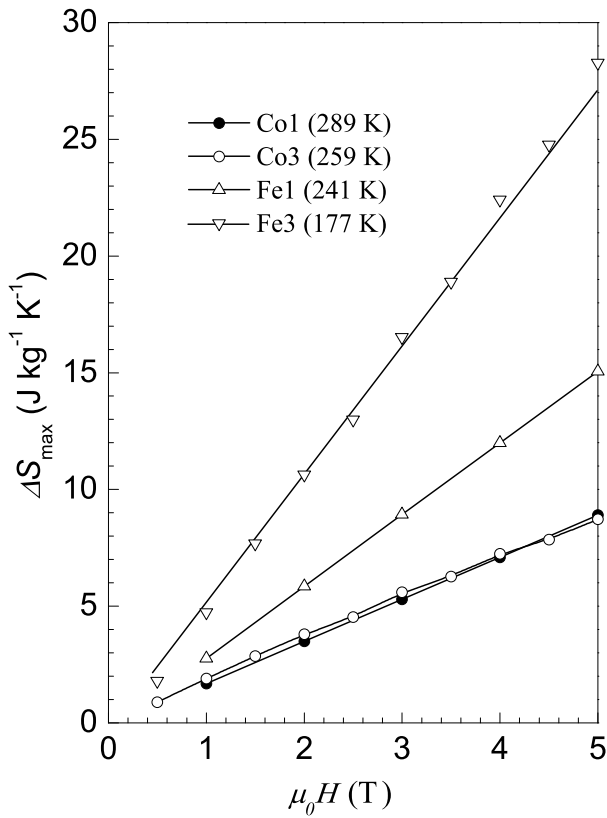}
\caption{\label{MCEmax} The maximum in the entropy change as a
function of applied field. The temperatures in the legend
correspond to the positions of the maxima in $\Delta S$ in Fig.
\ref{MCE}.}
\end{figure}

The temperature dependence of the entropy change for the presently
investigated samples are collected in Fig. \ref{MCE}. The addition
of either Fe or Co causes $e/a$ to decrease and, therefore, lowers
the temperature corresponding to the peak position of $\Delta S$.
There is not much change in the value of $\Delta S$ when Co is
alloyed. However, alloying 3 at\% Fe leads to a marked increase up
to 30 Jkg$^{-1}$K$^{-1}$ in $\Delta S$ for a field of 5T with
respect to the value of 20 Jkg$^{-1}$K$^{-1}$ for the non-alloyed
sample \cite{Krenke05a}. This increase is evidently related to a
comparatively large $\partial M/\partial T$ which would lead to a
large $\Delta S$ through Eq. 1. For this particular concentration,
a large $\partial M/\partial T$ can be found to lie in the large
spacing of the magnetization isotherms in Fig. 5b. Two additional
peaks in $\Delta S$ appear for the Co3 sample around 240 K and 260
K. These are expected to be related to the step-like features
observed below $M_s$ in the $M(T)$ data (Fig. 2b). Such features
can be associated with intermartensitic transformations, whereby
the modulation of the martensite structure changes.

The maximum entropy change $\Delta S_{max}$ obtained from Fig. 6
is plotted against the magnetic field in Fig. 7. In all cases, the
variation of $\Delta S_{max}$ with applied field is, to a good
approximation, linear within the range of the experiments. The
rate of increase of $\Delta S_{max}$ with field are 1.8 J
kg$^{-1}$ K$^{-1}$ T$^{-1}$ for Co1 and Co3; 3.0 J kg$^{-1}$
K$^{-1}$ T$^{-1}$ for Fe1 and 5.5 J kg$^{-1}$ K$^{-1}$ T$^{-1}$
for Fe3. There is no tendency to saturation for $\Delta S_{max}$
within the studied field range.

\section{Conclusion}

We have studied the effects of varying the electron concentration
by introducing small amounts of Fe and Co in place of Ni on the
magnetocaloric effect of Ni$_{50}$Mn$_{37}$Sn$_{13}$. Introducing
either of the elements leads to a decrease in $M_s$, whereby Co
reduces $\Delta S$, and 3 at\% Fe leads to an increase in $\Delta
S$. The increase is related to the increase in $\partial M/
\partial T$. The thermal hysteresis associated with the transition
becomes narrower with Co substitution, whereas it broadens when Fe
is added.

\begin{acknowledgments}
We thank Peter Hinkel for technical support. This work was
supported by Deutsche Forschungsgemeinschaft (GK277) and CICyT
(Spain), project MAT2004-1291. XM acknowledges support from DGICyT
(Spain).
\end{acknowledgments}


\end{document}